\documentclass[12pt]{article}
\usepackage[utf8]{inputenc}
\usepackage[british]{babel}
\usepackage{cmap}
\usepackage{lmodern}

\usepackage{amssymb, amsmath, amsthm}
\usepackage[a4paper,top=25mm,bottom=25mm,left=25mm,right=25mm]{geometry}
\usepackage{ragged2e}

\usepackage{authblk} 
\usepackage{pifont}
\usepackage{graphicx}
\usepackage[dvipsnames,svgnames,table]{xcolor}
\usepackage[figuresright]{rotating}
\usepackage{xtab} 
\usepackage{longtable} 
\usepackage{multirow}
\usepackage{footnote}
\usepackage[stable]{footmisc}
\usepackage{chngpage} 
\usepackage{pdflscape} 
\usepackage{tocbibind} 

\usepackage{pgfplots}
\pgfplotsset{compat=1.18}
\pgfplotsset{every tick label/.append style={font=\footnotesize}}
\usepackage{setspace}

\makesavenoteenv{tabular}
\usepackage{tabularx}
\usepackage{booktabs}
\usepackage{threeparttable} 
\usepackage[referable]{threeparttablex} 
\newcolumntype{R}{>{\raggedleft\arraybackslash}X}
\newcolumntype{L}{>{\raggedright\arraybackslash}X}
\newcolumntype{C}{>{\centering\arraybackslash}X}
\newcolumntype{A}{>{\columncolor{gray!25}}C}
\newcolumntype{a}{>{\columncolor{gray!25}}c}

\newlength{\tablen}

\usepackage{dcolumn} 
\newcolumntype{.}{D{.}{.}{-1}}

\usepackage{tikz}
\usetikzlibrary{arrows, calc, matrix, patterns, positioning, trees}
\usepackage[semicolon]{natbib}
\usepackage[hyphens]{url}
\usepackage{hyperref} 
\hypersetup{
  colorlinks   = true,    		
  urlcolor     = blue,    		
  linkcolor    = blue,			
  citecolor    = ForestGreen,		  	
}
\usepackage{microtype}
\usepackage[justification=centerfirst]{caption} 

\usepackage[labelformat=simple]{subcaption}

\DeclareCaptionLabelFormat{parenthesis}{(#2)}
\makeatletter
\renewcommand\p@subfigure{\arabic{figure}.}
\makeatother

\DeclareCaptionLabelFormat{parenthesis}{(#2)}
\makeatletter
\renewcommand\p@subtable{\arabic{table}.}
\makeatother

\usepackage{alphalph}

\def\addlegendimage{\csname pgfplots@addlegendimage\endcsname}

\usepackage{enumitem}

\setlist[itemize]{leftmargin=2.5\parindent}
\setlist[enumerate]{leftmargin=2.5\parindent}

\theoremstyle{plain}

\theoremstyle{definition}

\newtheorem{definition}{Definition}[section]
\newtheorem{example}{Example}

\theoremstyle{remark}

\newtheorem{remark}{Remark}

\def\keywords{\vspace{.5em} 
{\noindent \textit{Keywords}: }}

\def\AMS{\vspace{.5em} 
{\noindent \textbf{\emph{MSC} class}: }}

\def\JEL{\vspace{.5em} 
{\noindent \textbf{\emph{JEL} classification number}: }}

\title{Voting power in the Council of the European Union: \\ A comprehensive sensitivity analysis\thanks{~The views expressed in this paper are those of the authors and do not necessarily reflect the views of the Magyar Nemzeti Bank (MNB).}}
\author{
\href{https://sites.google.com/view/doragretapetroczy}{D\'ora Gr\'eta Petr\'oczy}\thanks{~E-mail: \emph{apetroczy@metropolitan.hu}
\newline Central Bank of Hungary, Budapest, Hungary
\newline Budapest Metropolitan University, MNB Institute, Budapest, Hungary}
$\qquad \qquad$
\href{https://sites.google.com/view/laszlocsato}{L\'aszl\'o Csat\'o}\thanks{~Corresponding author \newline
E-mail: \emph{laszlo.csato@sztaki.hun-ren.hu}
\newline Institute for Computer Science and Control (SZTAKI), Hungarian Research Network (HUN-REN), Laboratory on Engineering and Management Intelligence, Research Group of Operations Research and Decision Systems, Budapest, Hungary
\newline Corvinus University of Budapest (BCE), Institute of Operations and Decision Sciences, Department of Operations Research and Actuarial Sciences, Budapest, Hungary}
}
\date{\today}

\def\Dedication{
{\noindent
``\emph{Ideally, every responsible human being should have equal power in a world assembly. This situation, however, can only arise in an assembly of nations of equal sizes, each enfranchized to the same degree. To award nations voting powers proportional to their populations would give the spokesmen for large nations too much power in relation to the numbers of their sponsors.}''\footnote{~Source: \citet[p.~56]{Penrose1946}.}
}}

\begin{document}

\newgeometry{top=25mm,bottom=25mm,left=25mm,right=25mm}
\maketitle
\thispagestyle{empty}
\Dedication

\begin{abstract}
\noindent
The Council of the European Union (EU) is one of the main decision-making bodies of the EU. A number of decisions require a qualified majority, the support of 55\% of the member states (currently 15) that represent at least 65\% of the total population. We investigate how the power distribution based on the Shapley--Shubik index and the proportion of winning coalitions change if these criteria are modified within reasonable bounds. The power of the two countries, with approximately 4\% of the total population each, is found to be almost flat. The decisiveness index decreases if the population criterion is above 68\% or the states criterion is at least 17. Some quota combinations contradict the principles of double majority. The proportion of winning coalitions can be increased from 13.2\% to 20.8\% (30.1\%) such that the maximal relative change in the Shapley--Shubik indices remains below 3.5\% (5.5\%). Our results are indispensable for evaluating any proposal to reform the qualified majority voting system.
\end{abstract}


\AMS{91A80, 91B12}

\JEL{C71, D72}

\clearpage
\restoregeometry

\section{Introduction} \label{Sec1}

The Council of the European Union is one of the two legislative bodies in the European Union (EU), besides the European Parliament. Since the EU Council is an explicitly intergovernmental institution where a national minister represents each country, the distribution of power within the Council has a fundamental impact on the member states: the power politics view can explain as much as 90\% of the budget shares \citep{KauppiWidgren2004}. Although the role of political power may be lower, it is certainly a significant driver of budgetary allocation \citep{ZaporozhetsGarcia-ValinasKurz2016}. Hence, political science cannot ignore power indices \citep{FelsenthalLeechListMachover2003}.

The distribution of power in any institution depends on its voting rules. While some decisions require unanimity among the Council members, all major treaties have shifted further policy areas from unanimity to qualified majority. The definition of qualified majority was subject to several reforms in the past, especially because new members should have been provided with voting weights, which usually led to the modification of thresholds, too \citep{LeBretonMonteroZaporozhets2012}.
Unsurprisingly, the qualified majority voting system of the EU has been analysed in the fields of operational research \citep{Freixas2004, AlgabaBilbaoFernandez2007, AlonsoMeijideBilbaoCasasMendezFernandez2009}, political science \citep{AleskerovAvciIakoubaTurem2002, BindseilHantke1997, LaneMaeland2000}, and social choice theory \citep{BramsAffuso1976, BramsAffuso1985, FelsenthalMachover2001}.

Currently, Article 16 of the Treaty on European Union states the following conditions for a qualified majority:
\begin{itemize}[label=$\bullet$]
\item
\emph{Population criterion}: the supporting countries should represent at least 65\% of the EU population;
\item
\emph{States criterion}: the supporting countries should represent at least 55\% of the EU member states.
\end{itemize}
In addition, any blocking minority needs to contain at least four member states.
This system is effective from 1 November 2014.

Since any change in the set of member states is guaranteed to change voting powers, the effects of Brexit \citep{Grech2021, Kleinowski2025, Koczy2021, MaaserMayer2023}, as well as other possible entries \citep{Kirsch2022} and exits \citep{PetroczyRogersKoczy2022} have been thoroughly analysed. \citet{BhattacherjeeSarkar2019} even suggest an optimal threshold by maximising the correlation and minimising the differences in inequality between the weights and various power indices. However, \citet{BhattacherjeeSarkar2019} consider only the population criterion, which strongly limits the validity of their results, as will turn out in Section~\ref{Sec4}.

Our novel contribution to the literature resides in investigating how the power distribution in the EU Council will change if the majority rule of 65-55\% is modified and both the population and states thresholds are taken into account. It is assumed that the proportion of the total population required to accept a proposal varies between 51\% and 80\%, while the number of states required lies between 11 (40.7\%) and 20 (74.1\%).
Our calculations are based on the Shapley--Shubik index. This has several game-theoretical and statistical interpretations \citep{KurzMaaserNapel2018, LaruelleValenciano2001, LaruelleValenciano2007, LaruelleValenciano2009}; for example, the Shapley--Shubik index can be regarded as the expected relative share of the country in a fixed prize available to the winning coalition \citep{FelsenthalMachover2004}. Further arguments are provided in Section~\ref{Sec31}.

This issue is interesting not only for the academic community.
The Conclave, an assembly of 50 European leaders, published its Europe 2040 report in March 2024 and concluded that ``\emph{the generalisation of qualified majority voting in the Council should be achieved by 2030-2040}'' \citep{Conclave2024}. This report has been endorsed by the Conclave Board with \emph{Jean-Pierre Bourguignon}, the President of the European Research Council (ERC) from 2014 to 2019, among its members.
The German government has recently taken several steps towards the wider use of qualified majority voting, too \citep{OSW2023}: a Franco-German working group made recommendations for the institutional reform of the EU on 18 September 2023, including the adjustment of the thresholds to 60-60\%. However, the working group has not examined the impact of their proposal---which will be presented in the following, among many other results.

Our main findings can be summarised as follows:
\begin{itemize}
\item
The voting power of a country comprising about 4\% of the total population (the Netherlands and Romania) is almost insensitive to both thresholds of qualified majority (Section~\ref{Sec41});
\item
The current proportion of winning coalitions (13.2\%) decreases if the population criterion is above 68\% or the states criterion is at least 17 (Section~\ref{Sec42});
\item
The decisiveness index---the power of the Council of the EU to act---can be improved only by lowering the states threshold if the level of inequality (the power of large countries) is not allowed to decline (Section~\ref{Sec42});
\item
The recently proposed 60-60\% rule, which requires the support of 17 member states, strongly favours small countries and substantially reduces the proportion of winning coalitions (Section~\ref{Sec42});
\item
Some quota combinations contradict the principles of double majority as the level of inequality is higher than implied by population shares (Section~\ref{Sec42});
\item
A rule with a higher decisiveness index than the current 13.2\% is disadvantageous for either the majority of the countries or the majority of the population (Section~\ref{Sec43});
\item
The decisiveness index can be increased to 20.8\% (30.1\%) such that the maximal relative change in the Shapley--Shubik indices remains below 3.5\% (5.5\%) (Section~\ref{Sec43}).
\end{itemize}
These results are indispensable for evaluating any proposal to reform the criteria of the qualified majority voting system used in the EU Council.

The paper has the following structure.
Section~\ref{Sec2} provides a concise overview of related papers.
Section~\ref{Sec3} introduces the basic concepts of our approach (the Shapley--Shubik index, the decisiveness index, and the Herfindahl--Hirschman index) that are used for the sensitivity analysis in Section~\ref{Sec4}. Finally, Section~\ref{Sec5} ends with concluding remarks.

\section{Related literature} \label{Sec2}

A possible reason for the lack of sensitivity analyses in the extant literature is that measuring voting power in a weighted voting system is non-trivial. For instance, consider a simple situation with two voters, $A$ and $B$, having weights of 2 and 4, respectively. If the decision threshold is 3, the second voter $B$ is a dictator and holds all power. On the other hand, if the decision threshold is 5, both voters are veto players and they have the same power.

Several power indices have been proposed to quantify the influence or power of a voter in similar situations; \citet{Napel2018} provides a comprehensive summary of the main approaches and results. According to our knowledge, the first suggestion was made by \citet{Penrose1946}. Nonetheless, his approach has gained more attention only after \citet{Banzhaf1964} and \citet{Coleman1971} have ``rediscovered'' it; thus, it is called the Penrose--Banzhaf--Coleman, or, simply, the Banzhaf index.
The other popular power measure is the Shapley--Shubik index \citep{ShapleyShubik1954}. A recent proposal is the Coleman--Shapley index \citep{CasajusHuettner2019, Haimanko2020}.
In addition, voting power can be quantified by essentially any solution concept of cooperative game theory, such as the nucleolus \citep{Schmeidler1969, ZaporozhetsGarcia-ValinasKurz2016}.

Voting power has received intensive attention from academic researchers since the pioneering work of \citet{BramsAffuso1976}, which demonstrates that Luxembourg was a null player (had no power) in the Council of Ministers in the predecessor of the European Union. A comprehensive overview of this topic is provided by \citet{HollerNurmi2013}. However, some contributions from the last decade are worth presenting.


\citet{BhattacherjeeSarkar2019} perform a sensitivity analysis with respect to the population quota in the qualified majority voting rule of the Council. Their approach is similar to our work in that it handles the winning threshold as a controllable parameter. Seven measures of voting power and two indices of inequality are considered. Using Pearson's correlation coefficient as a similarity index, the maximum similarity of voting power to the original population share is found if the population quota is 66\%. However, this result is clearly due to disregarding the second, states criterion. Therefore, our crucial contribution to \citet{BhattacherjeeSarkar2019} is accounting for both the population and states quotas in the sensitivity analysis.

Several studies have shown independently that Brexit has mainly favoured the large countries \citep{Gabor2020, Grech2021, Gollner2018, Kirsch2016, KirschSlomczynskiStolickiZyczkowski2018, Koczy2021, MercikRamsey2017, Szczypinska2018}.  
On the other hand, a further exit would harm the large and benefit the small countries due to the unchanged states criterion at 15 \citep{PetroczyRogersKoczy2022}. \citet{KirschSlomczynskiStolickiZyczkowski2018} use the normal approximation of the Banzhaf index in double-majority games to uncover that such non-monotonicity is in most cases inherent in a double-majority system; however, it is strongly exacerbated by the peculiarities of the population vector in the European Union.

\citet{Kirsch2022} investigates the impact of the accession of Montenegro, Turkey, and Ukraine on the current 27 member states. In this case, the 55\% states threshold will grow from 15 to 16. Consequently, the power of small countries would increase, while the power of large countries would decrease. Because of its substantial size, the accession of Turkey would tip the balance of power: France, Germany, Italy, Poland, and Spain would lose almost 20\% of their voting power.

\citet{Kleinowski2024} develops a non-random voting model for the qualified majority voting in the Council, which abandons the assumption that each coalition of players is equally likely. It takes the role of the European Commission as an agenda-setter into account and, thus, excludes some coalitions from the analysis. \citet{Kleinowski2025} examines the post-Brexit council in the same framework. The ability of the seven most populous countries to build a blocking coalition is shown to have increased substantially after Brexit.

\section{Methodology} \label{Sec3}

This section describes the tools used for the investigation of voting powers.
Section~\ref{Sec31} discusses our choice of benchmarks for voting power. The Shapley--Shubik index is introduced in Section~\ref{Sec32}, while aggregated indices of decisiveness and inequality are presented in Section~\ref{Sec33}. Section~\ref{Sec34} provides a simple illustrative example. Finally, Section~\ref{Sec35} describes the practical details of the sensitivity analysis.

\subsection{Identifying the right benchmarks for voting power} \label{Sec31}

The two most popular measures of voting power are the Banzhaf and Shapley--Shubik indices.
\citet{Straffin1977} suggests two simple probability models to characterise these measures, respectively. Let $p_i$ denote the probability that a decision is supported by voter $i$, for all $1 \leq i \leq n$.
Under the \emph{independence} assumption, the values $p_i$ are selected independently and randomly from the unit interval. Therefore, each voter votes in favour of any proposal with a probability of 50\% since the uniform distribution on the unit interval has an average of 1/2.
Under the \emph{homogeneity} assumption, a common value $p$ is selected independently and randomly from the unit interval, and $p_i = p$ for all $1 \leq i \leq n$. Parameter $p$ can be interpreted as the ``level of acceptability'': some proposals are equally attractive to all voters ($p$ is close to 1), others are unreasonable ($p$ is close to 0), while a further set of proposals are controversial ($p$ is around 0.5).

Consider the probability that voter $i$ will be \emph{pivotal}, in other words, it will make a difference in the outcome. This probability equals the Banzhaf index of voter $i$ under the independence assumption \citep[Theorem~1]{Straffin1977}, and the Shapley--Shubik index of voter $i$ under the homogeneity assumption \citep[Theorem~2]{Straffin1977}. In our opinion, the preferences of the members in the EU Council are similar rather than independent of each other: if a bill is supported by some countries, this increases the probability that other countries agree with them.

\citet{KurzMaaserNapel2018} generalise the framework of \citet{Straffin1977} in a model where each voter has single-peaked preferences over an interval of alternatives. The delegates vote according to the median preference of their constituencies, which can be regarded as the countries in our case, and the assembly of delegates uses weighted majority voting. The ideal points of voters' preferences have a continuous distribution and are \emph{positively} correlated within each constituency, but they are independent across the constituencies.
Then the probability that a delegate has the decisive vote in the assembly asymptotically approaches its Shapley--Shubik index, not its Banzhaf value \citep[Proposition~1]{KurzMaaserNapel2018}. Furthermore, since the influence of a voter on the position of their delegate is inversely proportional to the size of the constituency (the population of the country), the Shapley--Shubik values of the delegates should be proportional to the size of their constituency in order to guarantee that any two voters have the same influence on the decision of the assembly. Consequently, the Shapley--Shubik indices in the EU Council will be compared to the population shares of the countries.

\subsection{The Shapley--Shubik index} \label{Sec32}

Voting situations are usually modelled by a cooperative game with transferable utility where the voters are the players, and the value of any coalition is maximal if it can accept a proposal and minimal otherwise.

Let $N$ denote the set of players, and $S \subseteq N$ be a coalition. The cardinal of a set is denoted by the corresponding small letter, namely, the number of players in coalition $S$ is $\lvert S \rvert = s$ and the number of players is $\lvert N \rvert = n$, respectively. The value of any coalition is given by the characteristic function $v: 2^N \to \mathbb{R}$.

\begin{definition}
\emph{Simple voting game}:
A game $(N,v)$ is called a simple voting game if
\begin{itemize}
\item
the payoffs are binary: $v(S) \in \left\{ 0,1 \right\} $ for all $S \subseteq N$;
\item
the superset of any winning coalition is a winning coalition: $v(S) = 1$ and $S \subseteq T$ implies $v(T) = 1$;
\item
there is at least one losing and one winning coalition: $v(\emptyset) = 0$ and $v(N) = 1$.
\end{itemize}
\end{definition}

The power of the players in voting games can be quantified by power indices. One popular measure for this purpose is the Shapley--Shubik index.

\begin{definition}
\emph{Shapley--Shubik index}:
Let $(N,v)$ be a simple voting game.
The Shapley--Shubik index of player $i$ is
\[
\varphi_i(N,v) = \sum_{S  \subseteq N \setminus  \{i\}}  \frac{s!(n-s-1)!}{n!} \left[ v \big( S \cup \left\{ i \right\} \big) -v \left( S \right) \right].
\]
\end{definition}
Consider a random order of the players. The Shapley--Shubik index of a player is its marginal contribution to the coalition formed by the preceding players, averaged over the set of all the possible orders of the players.

Several voting games are defined such that each player has a weight and a coalition becomes winning if the sum of the weights assigned to its players exceeds a given value.
Let vector $\mathbf{w} = \left[ w_i \right] \in \mathbb{R}^n$ denote the set of weights and $q$ denote the decision threshold.

\begin{definition}
\emph{Weighted voting game}:
A game $\left( N,v = \left[ \mathbf{w},q \right] \right)$ is called a weighted voting game if for any coalition $S \subseteq N$:
\[
v(S)= \left\{ 
\begin{array}{cl}
1 & \text{if }   \sum_{j \in S} w_j \geq q \\ 
0 & \text{otherwise}.
\end{array}
\right.
\]
\end{definition}

Consequently, the Shapley--Shubik index of a player in a weighted voting game is the probability that it becomes pivotal if the sequence of the players is chosen randomly from the $n!$ possible orders.
 
The qualified majority voting system of the EU Council can be formalised as a combination of three weighted voting games \citep[p.~1248]{KurzNapel2016}. In particular, the associated characteristic function is $v = \left( v^{(1)} \land v^{(2)} \right) \lor v^{(3)}$ or, equivalently, $v = v^{(1)} \land \left( v^{(2)} \lor v^{(3)} \right)$, where $\land$ is the AND and $\lor$ is the OR logical operator, while
\begin{itemize}
\item
$v^{(1)} = \left[ \mathbf{w}^{(1)}, q^{(1)} \right]$ such that $w_i^{(1)}$ is the population of country $i$ for all $1 \leq i \leq 27$ and $q^{(1)}$ is 65\% of the total population in the EU (population criterion);
\item
$v^{(2)} = \left[ \mathbf{w}^{(2)}, q^{(2)} \right]$ such that $w_i^{(2)} = 1$ for all $1 \leq i \leq 27$ and $q^{(2)} = 15$ (states criterion);
\item
$v^{(3)} = \left[ \mathbf{w}^{(3)}, q^{(3)} \right]$ such that $w_i^{(3)} = 1$ for all $1 \leq i \leq 27$ and $q^{(3)} = 24$ (blocking minority rule).
\end{itemize}

\subsection{Quantifying decisiveness and inequality} \label{Sec33}

The decision ability of the European Union is measured by the proportion of winning coalitions. This ``power of a collectivity to act'' has been defined by \citet{Coleman1971} and \citet{Carreras2005}, and axiomatically characterised by \citet{BaruaChakravartyRoy2009}.

\begin{definition}
\emph{Decisiveness index}:
Let $\left( N,v \right)$ be a simple voting game.
The decisiveness index is the number of winning coalitions divided by the number of possible coalitions $2^n$.
\end{definition}

We do not claim that a higher decisiveness index is preferable, as the decision process might be less stable if the countries can be combined in more ways into winning coalitions. However, an agreement is probably easier to reach if the number of winning coalitions is higher because a given set of countries is less likely to have veto power on the decision.

Two measures are used to quantify the inequality among the countries.
The first is a widely used measure of market concentration, the \emph{Herfindahl--Hirschman index} \citep{Herfindahl1950, Hirschman1945, Hirschman1964}. 

\begin{definition}
\emph{Herfindahl--Hirschman index ($\mathit{HHI}$)}:
Let $\left( N,v \right)$ be a simple voting game, $\lvert N \rvert = n$ be the number of players, and $\mathbf{\varphi} = \left[\varphi_i\right] \in \mathbb{R}^n$ be the vector of Shapley--Shubik indices.
The Herfindahl--Hirschman index is
\[
\mathit{HHI}(\mathbf{\varphi}) = \sum_{i=1}^n \varphi_i^2.
\]
\end{definition}

The maximum of $\mathit{HHI}$ is 1, reached if and only if a dictator exists in the game. The minimum of $\mathit{HHI}$ is $1/n$, reached if and only if all players have the same power. Since the lower bound of $\mathit{HHI}$ depends on the number of players, it is usual to consider its normalised version that lies between 0 and 1 \citep{PetroczyCsato2021}.

\begin{definition}
\emph{Normalised Herfindahl--Hirschman index ($\mathit{HHI}^\ast$)}:
Let $\left( N,v \right)$ be a simple voting game, $\lvert N \rvert = n$ be the number of players, and $\mathbf{\varphi} = \left[ \varphi_i \right] \in \mathbb{R}^n$ be the vector of Shapley--Shubik indices.
The normalised Herfindahl--Hirschman index is
\[
\mathit{HHI}^\ast(\mathbf{\varphi}) = \frac{HHI(\mathbf{\varphi})-1/n}{1-1/n}.
\]
\end{definition}

Thus, $\mathit{HHI}^\ast$ equals zero if all member states have the same power in the Council of the European Union. For example, this would be the case if the population criterion is ignored or, equivalently, the threshold is decreased below the population share of the smallest state. On the other hand, the normalised Herfindahl--Hirschman index would be one if one country is a dictator, namely, any coalition is winning if it contains this particular country and losing otherwise.

Therefore, $\mathit{HHI}^\ast$ reasonably reflects the inequality among the member states. In addition, since the voting powers of the members vary only in their population (otherwise, they are equal), a higher Herfindahl--Hirschman index indicates that the power of the small countries is generally lower.

The Gini coefficient is an index originally used to measure the degree of inequality in the distribution of income or wealth. However, it is also appropriate to quantify inequality in the distribution of voting power \citep{LaruelleValenciano2004}.

\begin{definition}
\emph{Gini index}:
Let $\left( N,v \right)$ be a simple voting game, $\lvert N \rvert = n$ be the number of players, and $\mathbf{\varphi} = \left[ \varphi_i \right] \in \mathbb{R}^n$ be the vector of Shapley--Shubik indices in an increasing order, that is, $\varphi_i \leq \varphi_j$ if $i < j$.
The Gini index is
\[
G(\mathbf{\varphi}) = \frac{\sum_{i=1}^n \sum_{j=1}^n \lvert \varphi_i - \varphi_j \rvert}{2n \sum_{i=1}^n \varphi_i} = \frac{1}{n} \left[ n+1- 2 \left( \frac{\sum_{i=1}^n (n + 1 - i) \varphi_i}{\sum_{i=1}^n \varphi_i} \right) \right] = \frac{2 \sum_{i=1}^n i \varphi_i - (n+1)}{n}
\]
because $\sum_{i=1}^n \varphi_i = 1$.
\end{definition}

Analogous to the normalised Herfindahl--Hirschman index, the minimum of $G$ is zero if all voting powers are equal. The Gini index has a maximum of $(n-1)/n$, which is reached if and only if a dictator exists in the voting game.

A lower value of $\mathit{HHI}^\ast$ and Gini should not necessarily be favoured since the power distribution should reflect the inherent inequalities between the countries in terms of their population to some extent. In other words, zero is not the appropriate benchmark. Thus, $\mathit{HHI}^\ast$ and Gini are mainly used to represent the relative power of small and large member states by a single value, which can be easily understood by the policy-makers. This advantage would be lost by adopting, for instance, the Penrose square root as the benchmark.
Nonetheless, if the Herfindahl--Hirschman or Gini index exceeds the value implied by the population shares, then the power of the large countries seems to be excessive.

\subsection{An illustrative example} \label{Sec34}

Take the following simple qualified majority voting system.

\begin{example} \label{Examp1}
A committee consists of three members $A$, $B$, and $C$, who have the voting weights 4, 2, and 1, respectively.
\end{example}

If the decision threshold in Example~\ref{Examp1} is $q=4$, then voter $A$ is a dictator and has the maximal power of 1. Even though the ratio of the voting weights is 4:2:1, the power distribution is 1:0:0.

\begin{table}[!t]
\caption{The sensitivity analysis of Example~\ref{Examp1} with respect to the quota $q$}
\label{Table1}
    \centering
    \rowcolors{1}{gray!20}{}
    \begin{tabularx}{\textwidth}{lCCCcCC} \toprule \hiderowcolors
    \multirow{2}[0]{*}{Threshold ($q$)} & \multicolumn{3}{c}{Shapley--Shubik index} & Decisiveness & \multirow{2}[0]{*}{$\mathit{HHI}^\ast$} & \multirow{2}[0]{*}{Gini} \\
          & $A$     & $B$     & $C$     & index &       &  \\ \bottomrule \showrowcolors
    3     &  2/3  &  1/6  &  1/6  & 62.5\% &  1/4  &  1/3 \\
    4     & 1     & 0     & 0     & 50\%  & 1     & 2/3     \\
    5     &  2/3  &  1/6  &  1/6  & 37.5\% &  1/4  &  1/3 \\
    6     &  1/2  &  1/2  & 0     & 25\%  &  1/4  &  1/3 \\
    7     &  1/3  &  1/3  &  1/3  & 12.5\% &  0  & 0     \\ \toprule
    \end{tabularx}
\end{table}

\begin{remark} \label{Rem1}
Consider Example~\ref{Examp1}.
The Shapley--Shubik indices of the voters are presented in Table~\ref{Table1} as a function of the majority threshold $q$.
While voter $A$ is a dictator and has maximal power if $q = 4$, this is not the case for any other value of $q$. Voter $C$ has no power if $q$ is odd, however, the powers of voters $B$ and $C$ are equal if $q$ is even.
\end{remark}


\begin{remark} \label{Rem2}
Consider Example~\ref{Examp1}.
There are three voters; hence, the number of coalitions is $2^3=8$. If the threshold is $q = 4$, then four winning coalitions exist ($A$, $\{ A, B \}$, $\{ A, C \}$, $\{ A, B, C \}$), and the decisiveness index is $4/8 = 50\%$.
Table~\ref{Table1} shows the proportion of winning coalitions for other values of $q$.
Since Example~\ref{Examp1} contains only one majority threshold $q$, the decisiveness index is a monotonically decreasing function of this parameter, as expected.
\end{remark}

\begin{remark} \label{Rem3}
Consider Example~\ref{Examp1}.
The Herfindahl--Hirschman and Gini indices are presented in Table~\ref{Table1} for various values of $q$.
The power distribution has the maximal level of inequality if $q = 4$, however, all voters have equal power if $q=7$. The Gini index equals $2/3$, its theoretical maximum, if $q=4$.
Note that the values of $\mathit{HHI}^\ast$ and $G$ remain the same if the threshold is increased from 5 to 6, even though the power of each voter changes.
\end{remark}

\subsection{Implementation} \label{Sec35}

In our sensitivity analysis, the Shapley--Shubik indices are calculated for all 27 current member states. The population criterion is investigated between 51\% and 80\%. 51\% is a natural lower bound, while it does not make sense to choose a quota over 80\% since Germany accounts for 18.59\% of the total population: in the absence of the specific blocking minority rule, the largest country would be a veto player for a threshold of at least 82\%.
The number of member states required for a qualified majority is assumed to be between 11 (which is 40.7\% of the number of EU countries) and 20 (74.1\%). A lower limit would be harmful for the small states, and a higher limit would make it challenging to accept any proposal.

\begin{table}[t!]
\caption{Populations and voting powers in the European Union}
\label{Table2}
    \centering
    \begin{threeparttable}     
    \rowcolors{1}{}{gray!20}
    \begin{tabularx}{0.9\textwidth}{lcCc} \toprule
    Country & Abbreviation & Population share & Shapley--Shubik index \\ \bottomrule
    Germany & DE    & 18.59\% & 17.87\% \\
    France & FR    & 15.16\% & 13.60\% \\
    Italy & IT    & 13.32\% & 11.69\% \\
    Spain & ES    & 10.60\% & \textcolor{white}{0}9.15\% \\
    Poland & PL    & \textcolor{gray!20}{0}8.41\% & \textcolor{gray!20}{0}6.84\% \\
    Romania & RO    & \textcolor{white}{0}4.25\% & \textcolor{white}{0}3.83\% \\
    Netherlands & NL    & \textcolor{gray!20}{0}3.96\% & \textcolor{gray!20}{0}3.61\% \\
    Belgium & BE    & \textcolor{white}{0}2.60\% & \textcolor{white}{0}2.62\% \\
    Greece & EL    & \textcolor{gray!20}{0}2.37\% & \textcolor{gray!20}{0}2.45\% \\
    Czech Republic & CZ    & \textcolor{white}{0}2.36\% & \textcolor{white}{0}2.44\% \\
    Sweden & SE    & \textcolor{gray!20}{0}2.33\% & \textcolor{gray!20}{0}2.43\% \\
    Portugal & PT    & \textcolor{white}{0}2.31\% & \textcolor{white}{0}2.41\% \\
    Hungary & HU    & \textcolor{gray!20}{0}2.16\% & \textcolor{gray!20}{0}2.31\% \\
    Austria & AT    & \textcolor{white}{0}2.00\% & \textcolor{white}{0}2.19\% \\
    Bulgaria & BG    & \textcolor{gray!20}{0}1.53\% & \textcolor{gray!20}{0}1.85\% \\
    Denmark & DK    & \textcolor{white}{0}1.31\% & \textcolor{white}{0}1.70\% \\
    Finland & FI    & \textcolor{gray!20}{0}1.24\% & \textcolor{gray!20}{0}1.64\% \\
    Slovakia & SK    & \textcolor{white}{0}1.21\% & \textcolor{white}{0}1.62\% \\
    Ireland & IE    & \textcolor{gray!20}{0}1.13\% & \textcolor{gray!20}{0}1.56\% \\
    Croatia & HR    & \textcolor{white}{0}0.86\% & \textcolor{white}{0}1.37\% \\
    Lithuania & LT    & \textcolor{gray!20}{0}0.63\% & \textcolor{gray!20}{0}1.19\% \\
    Slovenia & SI    & \textcolor{white}{0}0.47\% & \textcolor{white}{0}1.08\% \\
    Latvia & LV    & \textcolor{gray!20}{0}0.42\% & \textcolor{gray!20}{0}1.05\% \\
    Estonia & EE    & \textcolor{white}{0}0.30\% & \textcolor{white}{0}0.96\% \\
    Cyprus & CY    & \textcolor{gray!20}{0}0.20\% & \textcolor{gray!20}{0}0.89\% \\
    Luxembourg & LU    & \textcolor{white}{0}0.14\% & \textcolor{white}{0}0.85\% \\
    Malta & MT    & \textcolor{gray!20}{0}0.12\% & \textcolor{gray!20}{0}0.83\% \\ \toprule
    \end{tabularx}
    \begin{tablenotes} \footnotesize
        \item \emph{Notes}: Abbreviations are ISO 3166-1 alpha-2 codes.
        \item Population shares come from the Council Decision (EU, Euratom) 2022/2518 \citep{CouncilEuropeanUnion2022}, Shapley--Shubik indices are our own calculations.
    \end{tablenotes}
    \end{threeparttable}
\end{table}

The population shares of the countries, which determine their weights with respect to the population criterion, are updated every year. We use the data from the Council Decision (EU, Euratom) 2022/2518 of 13 December 2022 amending the Council's Rules of Procedure \citep[Annex~III]{CouncilEuropeanUnion2022}. The populations of the countries are divided by 10 thousand and rounded to the nearest integer. The population threshold is determined analogously. The Shapley--Shubik indices are computed by the python package \emph{powerindices} of \emph{Frank Huettner} \citep{Huettner2023}. These are exact values, not approximations by simulations.
The populations and the power distribution---the latter according to the current qualified majority system---are presented in Table~\ref{Table2}.

The blocking minority rule that requires at least four countries for the veto is disregarded. It has a marginal role as the proportion of blocking coalitions always remains above 50\% (see Section~\ref{Sec42}), that is, at least $2^{27}/2 = 67{,}108{,}864$ blocking coalitions exist, from which only 21 are excluded by the blocking minority rule. Furthermore, we mainly focus on the differences compared to the current power indices.

Ignoring the blocking minority rule is standard in the literature, see, for example, \citet[p.~404]{Kirsch2022}: ``\emph{The effect of the aforementioned ``additional rule,'' i.e., forbidding blocking decisions by only three states, is negligible in the present context. For the current EU, there are only 22 coalitions of 24 states which do not represent 65\% of the total population of the EU. This number is of (almost) no consequence for the power indices.}''
Obviously, the number of coalitions affected by the blocking minority rule depends on population shares of the countries; while there are 21 such coalitions based on the 2022 data, this increases to 22 based on the 2023 data, as well as based on the data used in \citet{Kirsch2022}.
\citet[Appendix~E]{PetroczyRogersKoczy2022} examine the effects of the blocking minority rule on the Shapley--Shubik indices and conclude that these are equivalent to ``rounding errors''.

\section{Results} \label{Sec4}

The effects of changing the qualified majority thresholds are analysed in three parts: Section~\ref{Sec41} discusses Shapley--Shubik indices at the level of countries, Section~\ref{Sec42} focuses on the effects of a reform on the whole European Union, and Section~\ref{Sec43} compares different population and states quota pairs with respect to the winners and losers of the change.

\subsection{Country-level changes in voting power} \label{Sec41}

\input{Figure1_voting_power_selected_countries}

In the current double majority system, the voting powers of two member states with approximately the same population are close; hence, the dynamics of their Shapley--Shubik indices are nearly identical. For example, Cyprus, Luxembourg, and Malta are similar, as well as Belgium, Greece, and Hungary.
Consequently, six different patterns can be identified that are plotted in Figure~\ref{Fig1}.

Germany, the largest state, almost always benefits from a higher population criterion and a lower states criterion. Its current voting power could not decrease with a states quota of 11, but could not increase with a states quota of 20. Although its power is now slightly below its share of the total population, this proportion is not impossible to exceed, especially for a lower states threshold. The curves for the next two largest countries, France and Italy, are similar.

Poland is also a large state, however, its voting power remains substantially below its population share at the moment.
The Shapley--Shubik index presents an unexpected function of the population quota:
(1) it is flat, then increasing, then decreasing if 11 states are required for a majority;
(2) it is increasing, then flat, then increasing, then decreasing if 15 or 17 states are required for a majority.
The power of Poland is higher for a lower states criterion, but it rarely achieves its population share, which could be unfair according to the main result of \citet{KurzMaaserNapel2018} that is discussed in Section~\ref{Sec31}.
The Shapley--Shubik index of Spain also has a maximum around 70--75\% population quota for any values of the states criterion between 11 and 20.

These peculiarities, especially the non-monotonicity of voting power as a function of the population quota, can probably be explained by the substitutability of the countries. The five highest differences for subsequent countries in the order of population shares appear between Poland and Romania, between Germany and France, between Italy and Spain, between Poland and Spain, and between France and Italy (each exceeding 1.5\%, see Table~\ref{Table2}). Consequently, these large countries, especially Poland, are the most difficult to replace with a country of similar size. On the other hand, the small countries can be easily substituted in a coalition, which essentially excludes any issues arising from discontinuity. Understanding these non-monotonicity results calls for future studies, similar to the ``paradox of redistribution'', when the voting power of a country decreases if its voting weight increases \citep{FischerSchotter1978, HollerNapel2004, HollerNapel2005}.

The voting power of Romania is remarkably robust with respect to both decision thresholds. Its Shapley--Shubik index can be slightly higher than its population share only for a high population criterion.

Belgium is a country whose voting power and population share currently nearly coincide. Its Shapley--Shubik index is slightly decreasing as the population threshold grows, except if only 11 states are required for a qualified majority. Its voting power is lower for a lower states criterion.

Croatia benefits from a lower population limit and a higher states threshold. Its voting power is sharply declining as a function of the population quota. Nonetheless, the power of Croatia is almost guaranteed to be higher than its share of the total population.

Finally, Malta is the smallest member state, hence, increasing the states criterion strongly favours it. The shapes of the functions are essentially analogous to the case of Croatia, although the gap between the voting power and the population share is naturally higher.

The Shapley--Shubik indices of the remaining 21 member states are not presented since their pattern are not fundamentally different from those of the six presented above.


\subsection{Aggregated changes: decisiveness index and inequality} \label{Sec42}

\begin{figure}[t!]
\centering

\begin{tikzpicture}
\begin{axis}[
xlabel = Population quota (\%),
x label style = {font=\small},
ylabel = Decisiveness index (\%),
y label style = {font=\small},
width = 0.99\textwidth,
height = 0.6\textwidth,
ymajorgrids = true,
xmin = 51,
xmax = 80,
ymin = 0,
legend style = {font=\small,at={(-0.05,-0.15)},anchor=north west,legend columns=5},
legend entries = {11 states$\qquad$,12 states$\qquad$,13 states$\qquad$,14 states$\qquad$,15 states,16 states$\qquad$,17 states$\qquad$,18 states$\qquad$,19 states$\qquad$,20 states}
] 
\draw [black,very thick,dashed] (\pgfkeysvalueof{/pgfplots/xmin},13.2110081613063) -- (\pgfkeysvalueof{/pgfplots/xmax},13.2110081613063);
\addplot [red, thick, dashdotdotted, mark=pentagon, mark size=2pt, mark options=solid] coordinates {
(51,46.3734664022922)
(52,44.1970810294151)
(53,42.0138075947761)
(54,39.83800932765)
(55,37.665044516325)
(56,35.511290282011)
(57,33.3755001425743)
(58,31.277409940958)
(59,29.2183630168437)
(60,27.2182680666446)
(61,25.2785600721836)
(62,23.4157159924507)
(63,21.6265365481376)
(64,19.9204437434673)
(65,18.2885572314262)
(66,16.7344786226749)
(67,15.2472779154777)
(68,13.8304024934768)
(69,12.4771393835544)
(70,11.1957371234893)
(71,9.98507514595985)
(72,8.8564246892929)
(73,7.81041234731674)
(74,6.85599073767662)
(75,5.99082633852958)
(76,5.21745979785919)
(77,4.52733263373374)
(78,3.91628593206405)
(79,3.37195321917533)
(80,2.88700759410858)
};
\addplot [gray, thick, mark=o, , mark size=2pt, mark options=solid] coordinates {
(51,44.2594483494758)
(52,42.2824189066886)
(53,40.2892284095287)
(54,38.2941357791423)
(55,36.2937107682228)
(56,34.3024641275405)
(57,32.3179699480533)
(58,30.3572103381156)
(59,28.4216411411762)
(60,26.5305593609809)
(61,24.6864639222621)
(62,22.9065097868442)
(63,21.1900293827056)
(64,19.5486851036548)
(65,17.9759912192821)
(66,16.4760991930961)
(67,15.0382257997989)
(68,13.6646375060081)
(69,12.3475670814514)
(70,11.09454408288)
(71,9.90516319870948)
(72,8.79225879907608)
(73,7.75885060429573)
(74,6.81532770395278)
(75,5.96018731594085)
(76,5.19583746790885)
(77,4.51339855790138)
(78,3.90822142362594)
(79,3.3678576350212)
(80,2.88517773151397)
};
\addplot [blue, thick, dashed, mark=asterisk, mark size=2.5pt, mark options={solid,semithick}] coordinates {
(51,40.2480110526084)
(52,38.5939829051494)
(53,36.9112841784954)
(54,35.2136574685573)
(55,33.5000202059745)
(56,31.7840240895748)
(57,30.0637364387512)
(58,28.3522605895996)
(59,26.6490630805492)
(60,24.9705515801906)
(61,23.3195766806602)
(62,21.7123374342918)
(63,20.1498843729496)
(64,18.6454057693481)
(65,17.1965263783931)
(66,15.809790790081)
(67,14.4766747951507)
(68,13.1991066038608)
(69,11.9686387479305)
(70,10.7903972268104)
(71,9.66250970959663)
(72,8.59762877225875)
(73,7.60091692209243)
(74,6.68637230992317)
(75,5.85593804717063)
(76,5.1140658557415)
(77,4.45252433419227)
(78,3.86612713336944)
(79,3.34132462739944)
(80,2.8701901435852)
};
\addplot [orange, thick, dashdotted, mark=square, mark size=2pt, mark options={solid,thin}] coordinates {
(51,34.0473473072052)
(52,32.8150190412998)
(53,31.543443351984)
(54,30.2445009350776)
(55,28.9185352623462)
(56,27.5782085955142)
(57,26.2237660586833)
(58,24.8662665486335)
(59,23.5037721693515)
(60,22.1477061510086)
(61,20.7990571856498)
(62,19.4715030491352)
(63,18.1660369038581)
(64,16.8947033584117)
(65,15.6578123569488)
(66,14.4645318388938)
(67,13.310495018959)
(68,12.1991671621799)
(69,11.1233107745647)
(70,10.0863724946975)
(71,9.08453464508056)
(72,8.12670513987541)
(73,7.21718296408653)
(74,6.3710793852806)
(75,5.59510812163353)
(76,4.89894077181816)
(77,4.27868366241455)
(78,3.73113676905632)
(79,3.24277058243751)
(80,2.8037391602993)
};
\addplot [black, thick, dotted, mark=star, mark size=2.5pt, mark options=solid] coordinates {
(51,26.1757731437683)
(52,25.3837794065475)
(53,24.5512574911117)
(54,23.685397952795)
(55,22.7872520685195)
(56,21.865852177143)
(57,20.9230624139308)
(58,19.9685156345367)
(59,19.0021827816963)
(60,18.031220883131)
(61,17.0550033450126)
(62,16.0819143056869)
(63,15.1126332581043)
(64,14.1555532813072)
(65,13.2110081613063)
(66,12.2874327003955)
(67,11.3843455910682)
(68,10.5068713426589)
(69,9.65099185705185)
(70,8.81975293159484)
(71,8.00964385271072)
(72,7.22585991024971)
(73,6.4694158732891)
(74,5.75145706534385)
(75,5.07955774664878)
(76,4.4668324291706)
(77,3.91652584075927)
(78,3.43120470643043)
(79,3.00203114748001)
(80,2.61992588639259)
};
\addplot [brown, thick, dashdotted, mark=diamond*, mark size=2pt, mark options=solid] coordinates {
(51,17.9257802665233)
(52,17.4982570111751)
(53,17.0398108661174)
(54,16.5523581206798)
(55,16.0355791449546)
(56,15.4946558177471)
(57,14.9309501051902)
(58,14.3510952591896)
(59,13.7566447257995)
(60,13.1535373628139)
(61,12.5410109758377)
(62,11.9236014783382)
(63,11.30061596632)
(64,10.6770530343055)
(65,10.052378475666)
(66,9.4312198460102)
(67,8.81367474794387)
(68,8.20503681898117)
(69,7.60405212640762)
(70,7.01408833265304)
(71,6.43301829695701)
(72,5.86452335119247)
(73,5.30814602971077)
(74,4.7699436545372)
(75,4.25366014242172)
(76,3.770212829113)
(77,3.32574844360351)
(78,2.92852148413658)
(79,2.57737040519714)
(80,2.26903706789016)
};
\addplot [purple, mark=oplus, mark size=2pt, mark options=solid] coordinates {
(51,10.7604190707206)
(52,10.5700954794883)
(53,10.3630684316158)
(54,10.1380757987499)
(55,9.89297479391098)
(56,9.62967425584793)
(57,9.3485340476036)
(58,9.05283167958259)
(59,8.74363258481025)
(60,8.42522606253624)
(61,8.09826105833053)
(62,7.76543617248535)
(63,7.42571800947189)
(64,7.08147585391998)
(65,6.73197358846664)
(66,6.37917146086692)
(67,6.02200329303741)
(68,5.66304549574852)
(69,5.30235543847084)
(70,4.94269877672195)
(71,4.5831486582756)
(72,4.2263463139534)
(73,3.87220159173011)
(74,3.52426096796989)
(75,3.18356975913047)
(76,2.85551771521568)
(77,2.54419967532157)
(78,2.25751996040344)
(79,1.99902802705764)
(80,1.77160799503326)
};
\addplot [ForestGreen, thick, dashed, mark=triangle*, mark size=2pt, mark options=solid] coordinates {
(51,5.58933168649673)
(52,5.52003011107444)
(53,5.44451624155044)
(54,5.36152198910713)
(55,5.26893734931945)
(56,5.16610667109489)
(57,5.0528660416603)
(58,4.93007153272628)
(59,4.79809716343879)
(60,4.65889871120452)
(61,4.51340600848197)
(62,4.36347350478172)
(63,4.20907661318779)
(64,4.05067950487136)
(65,3.88821512460708)
(66,3.72216925024986)
(67,3.55184152722358)
(68,3.37734892964363)
(69,3.19806486368179)
(70,3.0153751373291)
(71,2.82933339476585)
(72,2.64099016785621)
(73,2.45061591267585)
(74,2.2602155804634)
(75,2.07060500979423)
(76,1.88409388065338)
(77,1.70176029205322)
(78,1.52774155139923)
(79,1.36516094207763)
(80,1.21828317642211)
};
\addplot [violet, thick, loosely dashed, mark=otimes, mark size=2pt, mark options=solid] coordinates {
(51,2.48433127999305)
(52,2.46390104293823)
(53,2.4417333304882)
(54,2.41752117872238)
(55,2.39038243889808)
(56,2.35941857099533)
(57,2.32401937246322)
(58,2.28400155901908)
(59,2.23947092890739)
(60,2.19081342220306)
(61,2.13846042752265)
(62,2.08335667848587)
(63,2.02581658959388)
(64,1.96622833609581)
(65,1.90437287092208)
(66,1.84045583009719)
(67,1.77447870373725)
(68,1.70619413256645)
(69,1.63476914167404)
(70,1.56011879444122)
(71,1.48208439350128)
(72,1.40122026205062)
(73,1.31743848323822)
(74,1.23142823576927)
(75,1.14398300647735)
(76,1.05639547109603)
(77,0.968896597623825)
(78,0.882846117019653)
(79,0.799210369586944)
(80,0.720638781785965)
};
\addplot [Goldenrod, thick, mark=Mercedes star, mark size=3pt, mark options=solid] coordinates {
(51,0.934106856584549)
(52,0.92942863702774)
(53,0.924257934093475)
(54,0.918702781200408)
(55,0.91254785656929)
(56,0.905629992485046)
(57,0.897467136383056)
(58,0.887789577245712)
(59,0.876390188932418)
(60,0.863388180732727)
(61,0.848739594221115)
(62,0.832796841859817)
(63,0.815726071596145)
(64,0.797861069440841)
(65,0.779072195291519)
(66,0.759407132863998)
(67,0.738792866468429)
(68,0.717517733573913)
(69,0.695165991783142)
(70,0.671449303627014)
(71,0.645931810140609)
(72,0.618699938058853)
(73,0.589586794376373)
(74,0.558830797672271)
(75,0.526409596204757)
(76,0.493122637271881)
(77,0.45914351940155)
(78,0.424998253583908)
(79,0.390771031379699)
(80,0.357306748628616)
};
\end{axis}
\end{tikzpicture}

\caption{Decisiveness as a function of the population and states criteria \\ \vspace{0.2cm}
\footnotesize{\emph{Note}: The dashed black line shows the current decision probability.}}
\label{Fig2}

\end{figure}


Figure~\ref{Fig2} shows how the decisiveness index depends on the two thresholds. The general relationship is obvious: a higher population or states criterion reduces the number of coalitions which are able to accept a proposal. On the other hand, it is far from trivial that retaining the current level of decisiveness excludes a population quota over 68\%, as well as a states quota above 16. The recently suggested 60-60\% rule decreases the proportion of winning coalitions to 8.4\%.

\input{Figure3_decision_probability_HHI_Gini}

Figure~\ref{Fig3} presents the connection between the proportion of winning coalitions and the level of inequality in power distribution, as the function of the two majority criteria.
Under a fixed states threshold, a higher level of decisiveness usually implies a lower inequality, except for a high population criterion and a low states criterion. Consequently, if large countries want to maintain the status quo inequality, the decisiveness of the EU can be improved only by choosing a lower quota in the number of states required for double majority.

Interestingly, if the states criterion is only 11 or 12, the normalised Herfindahl--Hirschman index reaches its maximum approximately at the current decisiveness index of 13.2\%. The Gini index is also close to its maximum around the current decisiveness index if the states threshold is 11 or 12. However, such a high level of inequality would be difficult to explain since it exceeds the value associated with the population shares of the countries (see the dotted cyan line), and the qualified majority rule aims to mitigate the power of large countries. Therefore, a high population threshold together with a low states threshold seems to be unacceptable.
The recently suggested 60-60\% rule reduces the value of $\mathit{HHI}^\ast$ (Gini) by more than two-thirds (more than 43\%) compared to the status quo (65-55\%) as it decreases the population threshold and increases the states criterion to 17.

According to Figure~\ref{Fig3}, these implications are insensitive to the method used to quantify inequality in the distribution of voting power.

\subsection{The optimal qualified majority criteria} \label{Sec43}

Any change in the qualified majority thresholds reallocates voting powers among the member states, thus, it will produce some winners and losers as well. Now we focus on this aspect of the problem to understand which reforms can enjoy more support from the countries.

\input{Figure4_quota_pair_winners_status_quo}

Figure~\ref{Fig4} uncovers the proportion of winners: the number of states with a higher voting power (Figure~\ref{Fig4a}) and their population shares (Figure~\ref{Fig4b}). The number of winners varies between 4 and 23, while their proportion of the total population remains between 20.03\% and 84.66\%. Unsurprisingly, it is almost impossible to favour both the majority of citizens and the majority of countries, although two extreme sets of quotas (79\% of the population and 19 member states; 80\% of the population and 20 member states) are beneficial for 23 countries representing 51.8\% of the people. To conclude, any change is generally good for either the numerous small member states without a majority of the total population, or for the few large countries where more than half of the citizens live.

However, some pairs of criteria seem to be dominated. For example, the qualified majority rule (63\%, 14\#) favours 7 states representing 75\% of the population, while the neighbouring rule (62\%, 14\#) is advantageous for 12 countries representing 78.54\% of the population. Therefore, the rule (62\%, 14\#) is more likely to be accepted instead of (63\%, 14\#). Analogously, more countries (11 versus 8) and people (84.66\% instead of 77.6\%) benefit from the rule (59\%, 13\#) than from the rule (60\%, 13\#).
Consequently, even though a rule with a higher decisiveness index is disadvantageous for either the majority of the countries or the population, the analysis of the beneficiaries reveals the advantage of some qualified majority thresholds. 

\input{Figure5_quota_pair_maximal_loss}

Even if many member states with a substantial share of the total population are favoured by a change, a country will strongly oppose this reform if it loses much voting power. Thus, examining the maximal relative loss is also important because it might reflect the level of resistance against the change. Figure~\ref{Fig5} presents these values such that the size of the dots is proportional to the maximal relative loss. Intuitively, quota pairs along the diagonal generate the smallest maximal losses as simultaneous increases or decreases in the population and states criteria compensate each other to a certain extent. 

\begin{table}[t!]
  \centering
  \caption{Maximal relative losses in the Shapley--Shubik indices compared to the current qualified majority rule (65\%, 15\#) with a decisiveness index of 13.2\%}
  \label{Table3}
  \rowcolors{1}{gray!20}{}
    \begin{tabularx}{\textwidth}{CcCCC} \toprule \hiderowcolors
    \multicolumn{2}{c}{Double majority criteria} & \multicolumn{2}{c}{Maximal loss} & Decisiveness \\
    Population & Number of states & Value & Country & index \\ \bottomrule \showrowcolors
    61\%  & 14    & 1.98\% & Spain & 20.8\% \\
    64\%  & 15    & 2.54\% & Poland & 14.2\% \\
    72\%  & 17    & 3.19\% & Austria & \textcolor{gray!20}{0}4.2\% \\
    57\%  & 13    & 3.20\% & Germany & 30.1\% \\
    67\%  & 16    & 3.21\% & Romania & \textcolor{gray!20}{0}8.8\% \\ \toprule
    \end{tabularx}
\end{table}

\begin{figure}[t!]
\centering

\begin{subfigure}{\textwidth}
\caption{Qualified majority rule (61\%, 14\#)}
\label{Fig6a}

\begin{tikzpicture}
\begin{axis}[
width=\textwidth, 
height=0.6\textwidth,
symbolic x coords={MT,LU,CY,EE,LV,SI,LT,HR,IE,SK,FI,DK,BG,AT,HU,PT,SE,CZ,EL,BE,NL,RO,PL,ES,IT,FR,DE},
xlabel = Country,
xlabel style={font=\small},
ylabel = Relative change in voting power (\%) \\,
ylabel style={font=\small},
x tick label style={color=white},
x tick label style={major tick length=0pt},
ybar stacked,
ymajorgrids,
xtick align = inside,
bar width=6pt,
enlarge x limits={abs=0.5cm},
]

\addplot [blue,fill] coordinates {
(MT,1.44404332129964)
(LU,1.52941176470589)
(CY,1.57126823793492)
(EE,1.67014613778707)
(LV,1.72248803827753)
(SI,1.85013876040703)
(LT,1.8425460636516)
(HR,1.46198830409356)
(IE,1.66666666666666)
(SK,1.66769610870907)
(FI,1.71149144254279)
(DK,1.23747790218034)
(BG,1.56756756756757)
(AT,1.77757520510482)
(HU,2.08242950108459)
(PT,2.19825798423889)
(SE,2.22680412371135)
(CZ,2.21039705280393)
(EL,2.2439820481436)
(BE,2.48565965583174)
(NL,3.32317917474383)
(RO,3.44917690096682)
(PL,0)
(ES,-1.97835829052356)
(IT,-1.30937098844672)
(FR,-1.47848473703567)
(DE,-1.94223665062129)
};

\node [rotate=90,left] at (axis cs:MT,0){MT};
\node [rotate=90,left] at (axis cs:LU,0){LU};
\node [rotate=90,left] at (axis cs:CY,0){CY};
\node [rotate=90,left] at (axis cs:EE,0){EE};
\node [rotate=90,left] at (axis cs:LV,0){LV};
\node [rotate=90,left] at (axis cs:SI,0){SI};
\node [rotate=90,left] at (axis cs:LT,0){LT};
\node [rotate=90,left] at (axis cs:HR,0){HR};
\node [rotate=90,left] at (axis cs:IE,0){IE};
\node [rotate=90,left] at (axis cs:SK,0){SK};
\node [rotate=90,left] at (axis cs:FI,0){FI};
\node [rotate=90,left] at (axis cs:DK,0){DK};
\node [rotate=90,left] at (axis cs:BG,0){BG};
\node [rotate=90,left] at (axis cs:AT,0){AT};
\node [rotate=90,left] at (axis cs:HU,0){HU};
\node [rotate=90,left] at (axis cs:PT,0){PT};
\node [rotate=90,left] at (axis cs:SE,0){SE};
\node [rotate=90,left] at (axis cs:CZ,0){CZ};
\node [rotate=90,left] at (axis cs:EL,0){EL};
\node [rotate=90,left] at (axis cs:BE,0){BE};
\node [rotate=90,left] at (axis cs:NL,0){NL};
\node [rotate=90,left] at (axis cs:RO,0){RO};
\node [rotate=90,left] at (axis cs:PL,0){PL};
\node [rotate=90,right] at (axis cs:ES,0){ES};
\node [rotate=90,right] at (axis cs:IT,0){IT};
\node [rotate=90,right] at (axis cs:FR,0){FR};
\node [rotate=90,right] at (axis cs:DE,0){DE};

\draw [ultra thick] ({rel axis cs:0,0}|-{axis cs:MT,0}) -- ({rel axis cs:1,0}|-{axis cs:DE,0});
\end{axis}
\end{tikzpicture}
\end{subfigure}

\vspace{0.25cm}
\begin{subfigure}{\textwidth}
\caption{Qualified majority rule (57\%, 13\#)}
\label{Fig6b}

\begin{tikzpicture}
\begin{axis}[
width=\textwidth, 
height=0.6\textwidth,
symbolic x coords={MT,LU,CY,EE,LV,SI,LT,HR,IE,SK,FI,DK,BG,AT,HU,PT,SE,CZ,EL,BE,NL,RO,PL,ES,IT,FR,DE},
xlabel = Country,
xlabel style={font=\small},
ylabel = Relative change in voting power (\%) \\,
ylabel style={font=\small},
x tick label style={color=white},
x tick label style={major tick length=0pt},
ybar stacked,
ymajorgrids,
xtick align = inside,
bar width=6pt,
enlarge x limits={abs=0.5cm},
]

\addplot [blue,fill] coordinates {
(MT,2.76774969915765)
(LU,2.70588235294116)
(CY,2.6936026936027)
(EE,2.81837160751566)
(LV,2.87081339712918)
(SI,2.86771507863088)
(LT,2.68006700167505)
(HR,2.04678362573101)
(IE,1.92307692307694)
(SK,1.91476219888822)
(FI,1.95599022004891)
(DK,1.35533294048322)
(BG,1.67567567567568)
(AT,1.64083865086599)
(HU,1.90889370932754)
(PT,1.94939858979677)
(SE,1.97938144329897)
(CZ,1.96479738027016)
(EL,1.99918400652794)
(BE,2.17973231357553)
(NL,2.68623649958459)
(RO,2.82205382806378)
(PL,5.3639286758258)
(ES,-1.82533610230626)
(IT,-1.54043645699614)
(FR,-2.57447591026111)
(DE,-3.19601477667079)
};

\node [rotate=90,left] at (axis cs:MT,0){MT};
\node [rotate=90,left] at (axis cs:LU,0){LU};
\node [rotate=90,left] at (axis cs:CY,0){CY};
\node [rotate=90,left] at (axis cs:EE,0){EE};
\node [rotate=90,left] at (axis cs:LV,0){LV};
\node [rotate=90,left] at (axis cs:SI,0){SI};
\node [rotate=90,left] at (axis cs:LT,0){LT};
\node [rotate=90,left] at (axis cs:HR,0){HR};
\node [rotate=90,left] at (axis cs:IE,0){IE};
\node [rotate=90,left] at (axis cs:SK,0){SK};
\node [rotate=90,left] at (axis cs:FI,0){FI};
\node [rotate=90,left] at (axis cs:DK,0){DK};
\node [rotate=90,left] at (axis cs:BG,0){BG};
\node [rotate=90,left] at (axis cs:AT,0){AT};
\node [rotate=90,left] at (axis cs:HU,0){HU};
\node [rotate=90,left] at (axis cs:PT,0){PT};
\node [rotate=90,left] at (axis cs:SE,0){SE};
\node [rotate=90,left] at (axis cs:CZ,0){CZ};
\node [rotate=90,left] at (axis cs:EL,0){EL};
\node [rotate=90,left] at (axis cs:BE,0){BE};
\node [rotate=90,left] at (axis cs:NL,0){NL};
\node [rotate=90,left] at (axis cs:RO,0){RO};
\node [rotate=90,left] at (axis cs:PL,0){PL};
\node [rotate=90,right] at (axis cs:ES,0){ES};
\node [rotate=90,right] at (axis cs:IT,0){IT};
\node [rotate=90,right] at (axis cs:FR,0){FR};
\node [rotate=90,right] at (axis cs:DE,0){DE};

\draw [ultra thick] ({rel axis cs:0,0}|-{axis cs:MT,0}) -- ({rel axis cs:1,0}|-{axis cs:DE,0});
\end{axis}
\end{tikzpicture}
\end{subfigure}

\caption{Relative changes in the Shapley--Shubik indices compared to the current qualified majority rule (65\%, 15\#) \\ \vspace{0.2cm}
\footnotesize{\emph{Note}: See Table~\ref{Table2} for the abbreviations of country names.}}
\label{Fig6}

\end{figure}
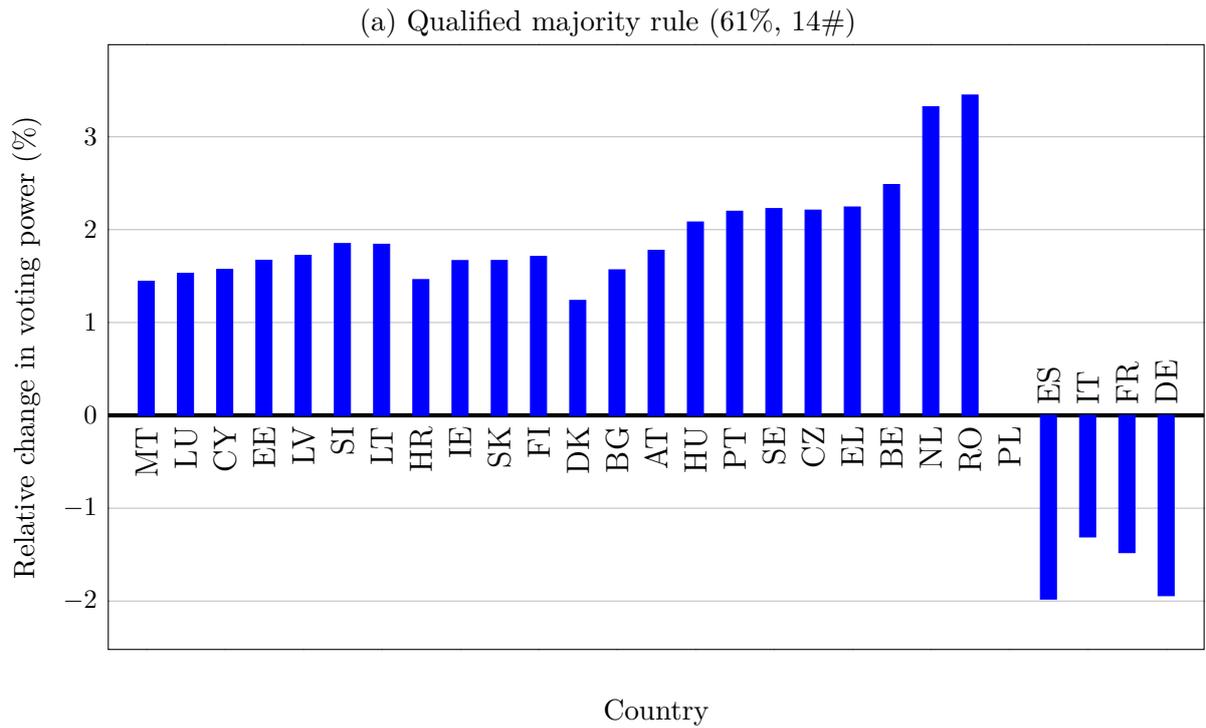
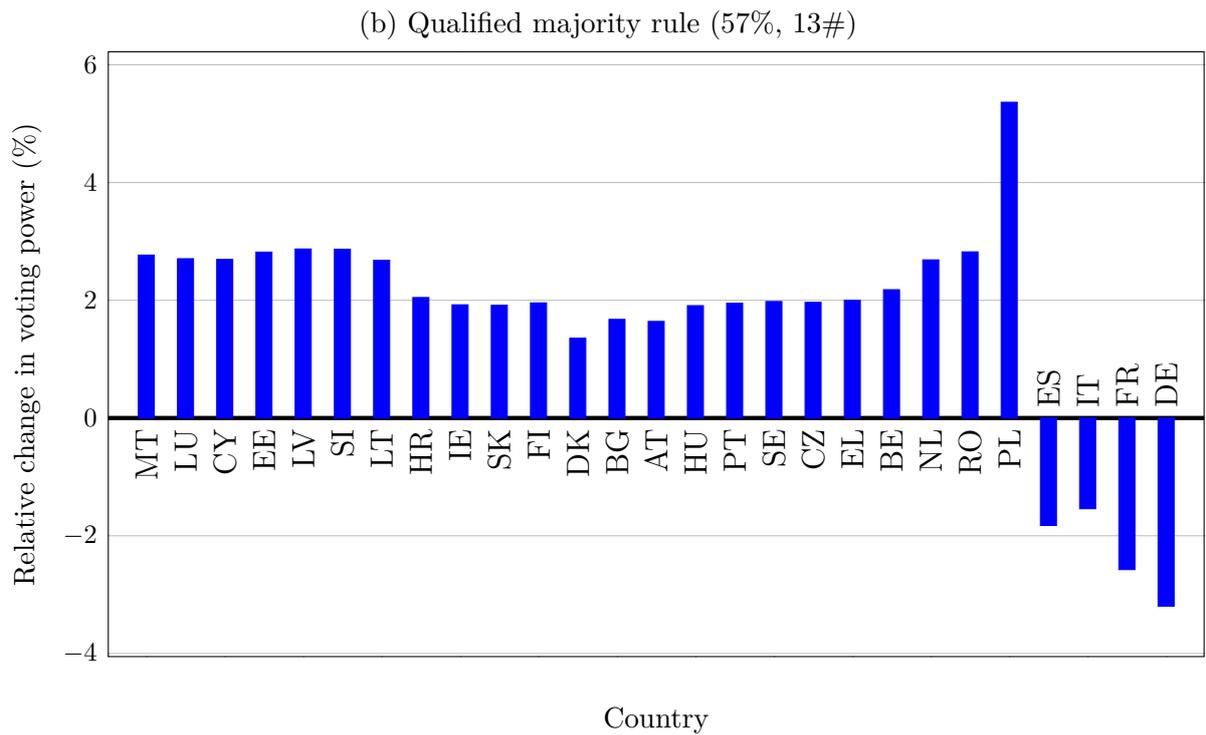


Table~\ref{Table3} reports the five minimal values. If the decisiveness index is not allowed to decrease, the rule (61\%, 14\#) seems to be the best option, followed by (57\%, 13\#), since (64\%, 15\#) is close to the status quo.
The implied changes in the voting powers are plotted in Figure~\ref{Fig6}. The pattern is similar in the two cases for all countries except for Poland, which gains nothing if the population threshold is reduced to 61\% and the states criterion to 14 (Figure~\ref{Fig6a}), but benefits the most from the alternative rule (57\%, 13\#) (Figure~\ref{Fig6b}). The reason is highlighted in Figure~\ref{Fig1}: around a population quota of 60\%, the voting power of Poland is insensitive to changing this criterion, while it is favoured by lowering the states criterion.
Crucially, the current decisiveness index of 13.2\% can be increased to 20.8\% such that the maximal change in the Shapley--Shubik indices remains below 3.5\%, or even to 30.1\% such that the maximal relative change in the Shapley--Shubik indices remains below 5.5\%.

\input{Figure7_quota_pair_winners_pop_share}

Finally, Figure~\ref{Fig7} considers gains and losses in the light of equal voting power for all citizens, which has been justified in Section~\ref{Sec31}. Here, the winners (losers) are the voters who have a higher (lower) power than the average voter. As can be seen in Table~\ref{Table2}, the current qualified majority system favours 20 member states in this respect (Figure~\ref{Fig7a}), but they account for only 25\% of the total population (Figure~\ref{Fig7b}). That is the predominant situation: for instance, if the states criterion remains 15, the population threshold can vary between 51\% and 74\% without changing the set of winners and losers. Unsurprisingly, the majority of the EU population usually has a lower than average voting power, while most countries benefit from the double majority rule. Some quota pairs favour more than 50\% percent of the member states and the total population---for example, the rule (76\%, 13\#)---but they have a high population threshold and imply low decisiveness, as well as high inequality. Therefore, our previous finding that a rule with a higher decisiveness index than the current 13.2\% is disadvantageous for either the majority of the countries or the majority of the population remains unchanged if the notion of fairness is changed to one person, one vote.

\section{Conclusions} \label{Sec5}

Since 2014, the Council of the European Union applies a complex rule of qualified majority: a decision requires the support of at least 55\% of the member states (currently 15 countries) that represent 65\% of the total population. We have investigated the effects of modifying these arbitrary thresholds with respect to the voting power of individual countries, the (in)equality of the power distribution, and the decisiveness index of the EU. According to our findings, even though any reform will be harmful to the majority of either the countries or the population, policy-makers can choose some quota pairs that minimise resistance and substantially increase (or decrease) the decisiveness of the EU.

Because the voting system of the Council of the European Union is able to produce unexpected and unwanted changes, similar analyses of power indices are important to understand and, possibly, to improve it. Hopefully, the current study will inspire further research along this line.
In particular, it would be instructive to look at another widely used power measure, the Banzhaf index. Although the Shapley–Shubik and Banzhaf indices have fairly similar values in many cases, they can behave quite differently in certain settings \citep[pp.~277--278]{FelsenthalMachover1998}.
Second, our paper provides only a snapshot of power distribution, but the population shares of the countries change dynamically. Hence, following \citet{Koczy2012}, it would be worth looking at the long-term effects of modifying the qualified majority voting system based on population predictions.
Third, the impact of expected EU enlargements can be examined \citep{Kirsch2022}.
Last but not least, the preferences of the countries to coalesce can also be taken into account; power indices to that end have been proposed by \citet{AleskerovHollerKamalova2014, BenatiVittucciMarzetti2013, BenatiVittucciMarzetti2021, PajalaWidgren2004}, among others.

\section*{Acknowledgements}
\addcontentsline{toc}{section}{Acknowledgements}
\noindent
Two anonymous reviewers and eight colleagues provided useful comments on earlier drafts. \\
We are grateful to \emph{Frank Huettner} for developing the python package \emph{powerindices}. \\
The research was supported by the National Research, Development and Innovation Office under Grants Advanced 152220, FK 145838, and PD 146055, by the J\'anos Bolyai Research Scholarship of the Hungarian Academy of Sciences, and by the EK\"OP-24 University Research Scholarship Program of the Ministry for Culture and Innovation from the source of the National Research, Development and Innovation Fund. \\
\emph{D\'ora Gr\'eta Petr\'oczy} was awarded a fellowship by the Hungarian Academy of Sciences to support researchers raising children.

\bibliographystyle{apalike}
\bibliography{All_references}

\end{document}